\documentclass[fleqn,12pt,twoside]{article}
\usepackage{espcrc1}

\usepackage{graphicx}
\usepackage[figuresright]{rotating}

\newcommand{\AmS}{{\protect\the\textfont2
  A\kern-.1667em\lower.5ex\hbox{M}\kern-.125emS}}

\title{An Infrared Renormalization Group Limit Cycle in QCD}

\author{Eric Braaten\address[MCSD]{Department of Physics,
         			The Ohio State University, 
				Columbus, OH\ 43210, USA}
       \thanks{Supported in part by DOE grant DE-FG02-91-ER4069.} 
        and
        H.-W.~Hammer\address{Universit{\"a}t Bonn, HISKP (Theorie), 
			Nussallee 14-16, 53115 Bonn, Germany}}
       
\begin{document}

\maketitle


The successes of the renormalization group (RG)
range from the explanation for universality in critical phenomena
to the nonperturbative formulation of quantum field theories.
The RG can be reduced to a set of differential equations 
that define a flow in the space of dimensionless coupling constants. 
Asymptotic {\it scale invariance} at long distances, as in critical phenomena, 
can be explained by RG flow to an {\it infrared fixed point}.
Asymptotic scale invariance at short distances in
quantum field theories 
can be explained by RG flow to an {\it ultraviolet fixed point}.
A fixed point is the simplest topological feature 
that can be exhibited by a RG flow.  As pointed out by Wilson in 1971,
the next simplest possibility is a {\it limit cycle}, 
a 1-dimensional orbit that is closed under the RG flow \cite{Wilson:1970ag}.
The RG flow carries the system 
around a complete orbit every time
the ultraviolet cutoff $\Lambda$ increases by some factor $\lambda_0$.
One of the signatures of an RG limit cycle
is {\it discrete scale invariance}: symmetry with respect to scaling
by a factor $\lambda_0^n$ only for integer values of $n$.  
Simple discrete Hamiltonian systems with RG limit cycles
have been constructed \cite{Glazek:2002hq,LeClair:2002ux}.
An ultraviolet limit cycle has been discovered in a 
familiar 2-dimensional field theory \cite{Leclair:2003xj}.
Ultraviolet limit cycles have also been found in 
effective field theories (EFT's) for bosons 
with large scattering length \cite{Bedaque:1998kgkm}
and for low-energy nucleons \cite{Bedaque:1999ve}.
In this talk, we point out that 
Quantum Chromodynamics (QCD) may have an infrared limit cycle
in the 3-nucleon sector for critical values 
of the up and down quark masses. 

In 1970, Efimov discovered that if the scattering length $a$ 
of nonrelativistic bosons is much larger than the 
range of the interaction, there are shallow 3-body bound states
whose number increases logarithmically with $|a|$ \cite{Efimov70}.
In the resonant limit $a \to \pm \infty$, 
there are infinitely-many arbitrarily shallow {\it Efimov states}.
The ratio of successive binding energies 
rapidly approaches a universal constant
$\lambda_0^2 = e^{2\pi/s_0} \approx 515$, where 
$s_0\approx 1.00624$ is a transcendental number.
The Efimov effect was revisited by Bedaque, Hammer, 
and van Kolck (BHvK) within the EFT framework \cite{Bedaque:1998kgkm}.
Bosons with large scattering length $a$ 
can be described by a nonrelativistic field theory 
with a 2-body contact interaction $(g_2/4) (\psi^*\psi)^2$.
In the 2-body sector, renormalization can be implemented 
by adjusting the coupling constant $g_2(\Lambda)$ 
as a function of the ultraviolet 
momentum cutoff $\Lambda$ so that the scattering length is $a$.
As $\Lambda \to \infty$, the dimensionless coupling constant 
$\Lambda g_2(\Lambda)$ flows to an ultraviolet fixed point:
$\Lambda g_2(\Lambda) \to - 4 \pi^2$.
The renormalization of $g_2$ eliminates ultraviolet divergences
from 3-body observables, but they still depend on 
the ultraviolet cutoff as
periodic functions of $\ln(\Lambda)$ with period $\pi/s_0$.
BHvK showed that the 3-body sector could be fully renormalized 
by including a 3-body contact interaction $(g_3/36)(\psi^* \psi)^3$ 
\cite{Bedaque:1998kgkm}.  As $\Lambda \to \infty$,  
the dimensionless coupling constant $\Lambda^4 g_3(\Lambda)$ 
flows to an ultraviolet limit cycle:
$\Lambda^4 g_3(\Lambda) \to - 144 \pi^4 H(\Lambda)$ , where
\begin{eqnarray}
H(\Lambda) &=& {\cos[s_0 \ln(\Lambda/\Lambda_*) + \arctan(s_0)] 
	\over \cos[s_0 \ln(\Lambda/\Lambda_*) - \arctan(s_0)]} 
\label{Hlc}
\end{eqnarray}
and $\Lambda_*$ is a scaling-violation parameter 
introduced by dimensional transmutation.
In the resonant limit $a \to \pm \infty$, there is an 
infrared limit cycle. The asymptotic discrete scaling 
symmetry with scaling factor $\lambda_0 \approx 22.7$ 
is manifest in the spectrum of Efimov states.

The spin-singlet and spin-triplet scattering lengths 
$a_s$ and $a_t$ for nucleons
are both significantly larger than the range of the nuclear force. 
Efimov used this observation as the basis for a qualitative approach 
to the few-nucleon problem in which the nuclear forces are approximated
by zero-range potentials with depths adjusted 
to reproduce the scattering lengths $a_s$ and $a_t$ \cite{Efi81}. 
This approach gives a reasonably accurate prediction 
for the deuteron binding energy.  In the 3-nucleon sector, 
the Efimov effect makes it necessary to impose a boundary condition 
on the wavefunction at short distances that can be fixed by using 
the triton binding energy $B_t$ as input.
The resulting prediction for
the neutron-deuteron scattering length is surprisingly accurate.
Efimov's program has been revisited by BHvK 
within the EFT framework \cite{Bedaque:1999ve}.
The EFT involves an isospin doublet $N$
of Pauli fields with two independent 2-body contact interactions:
$N^\dagger \sigma_i N^c N^{c \dagger} \sigma_i N$ 
and  $N^\dagger \tau_k N^c N^{c \dagger} \tau_k N$,
where $N^c = \sigma_2 \tau_2 N^*$.
Renormalization in the 2-body sector requires 
the two coupling constants to be adjusted as functions of 
$\Lambda$ to obtain the correct values of $a_s$ and $a_t$.
Renormalization in the 3-body sector requires 
a 3-body contact interaction 
$N^\dagger \sigma_i N^c N^{c \dagger} \sigma_j N 
	N^\dagger \sigma_i \sigma_j N$
with a coupling constant proportional to (\ref{Hlc}).
Thus the renormalization involves an ultraviolet limit cycle.
The scaling-violation parameter $\Lambda_*$ can be determined by
using the triton binding energy $B_t$ as input.

The low-energy few-nucleon problem can also be described 
by an EFT that includes explicit pion fields.
Such an EFT has been used to extrapolate nuclear forces 
to the chiral limit of QCD in which the pion is exactly 
massless \cite{Beane:2002xf,Epelbaum:2002gb}. 
The extrapolation to larger values of $m_\pi$ predicts
that $a_t$ diverges and 
the deuteron becomes unbound at a critical 
value in the range 170 MeV $< \, m_\pi \, <$ 210 MeV.
It is also predicted that $a_s$ is likely to diverge
and the spin-singlet deuteron become bound 
at some critical value of $m_\pi$ not much larger than 150 MeV. 
Both critical values are close to the physical value
$m_\pi = 135$ MeV.  

\begin{figure}[tb]
\centerline{\includegraphics*[width=9cm,angle=0,clip=true]{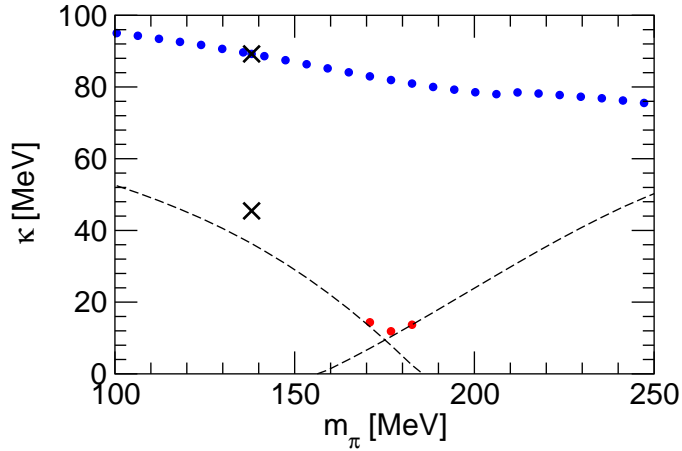}}
\caption{The binding momenta $\kappa=(mB_3)^{1/2}$ 
of $p n n$ bound states as a function of $m_\pi$.
The circles are the triton ground state and excited state. 
The crosses are the physical binding energies of the 
deuteron and triton.  The dashed lines are the thresholds
for decay into a nucleon plus a deuteron (left curve) or 
a spin-singlet deuteron (right curve).
}              
\label{fig:spec}
\end{figure}

Chiral extrapolations can also be calculated 
using the EFT without pions \cite{Braaten:2003eu}.
The inputs required are the chiral extrapolations $a_s(m_\pi)$, 
$a_t(m_\pi)$, and $B_t(m_\pi)$,
which can be calculated using an EFT with pions.
As an illustration, we take the central values of the error
bands for the inverse scattering lengths $1/a_s(m_\pi)$ and $1/a_t(m_\pi)$ 
from the chiral extrapolation in Ref.~\cite{Epelbaum:2002gb}.
Since the chiral extrapolation of the 
triton binding energy $B_t(m_\pi)$ has not yet been calculated and
since $\Lambda_*$ should vary smoothly with $m_\pi$,
we approximate it by its physical value $\Lambda_* =189$ MeV 
for $m_\pi=138$ MeV.
In Fig.~\ref{fig:spec}, we show the resulting 3-body spectrum in the triton 
channel as a function of $m_\pi$.  
Near $m_\pi \approx 175$ MeV where the decay threshold 
comes closest to $\kappa = 0$,
an excited state of the triton appears.
This excited state is a hint 
that the system is very close to an infrared limit cycle.
In the case illustrated by Fig.~\ref{fig:spec}, the value of $m_\pi$
at which $a_t$ diverges is larger than that at which $a_s$ diverges.
If they both diverged at the same value of $m_\pi$,
there would be an exact infrared limit cycle.

We conjecture that QCD can be tuned to this infrared limit cycle 
by adjusting the up and down quark masses $m_u$ and $m_d$.
As illustrated in Fig.~\ref{fig:spec}, the tuning of $m_\pi$,
which corresponds to $m_u + m_d$, is likely to bring 
the system close enough to the infrared limit cycle for the triton
to have one excited state.  We conjecture that by adjusting
the two parameters $m_u$ and $m_d$ to critical values,
one can make $a_t$ and $a_s$ diverge simultaneously.
At this critical point, the deuteron and spin-singlet deuteron would both 
have zero binding energy and the triton would have 
infinitely-many increasingly-shallow excited states.
The ratio of the binding energies of successively shallower states 
would rapidly approach a constant $\lambda_0^2$ close to 515.  
It may be possible to use a combination of lattice gauge theory and EFT 
to demonstrate the existence of this infrared RG limit cycle in QCD.

\end{document}